\documentclass[epj]{svjour}

\usepackage{graphics}
\usepackage{epsfig}
\usepackage[english]{babel}
\usepackage{euscript,amstext,amssymb,amsbsy,fancybox,epsf}

\title{Dispersion Theory and the Low Energy Constants for Neutral Pion Photoproduction}
\author{B. Pasquini\inst{1}, D. Drechsel\inst{2}, L. Tiator\inst{2}
}                     

\authorrunning{B. Pasquini et al.}%
\institute{Dipartimento di Fisica Nucleare e Teorica, Universit\`a degli Studi di Pavia;
INFN, Sezione di Pavia, Pavia, Italy \and Institut f\"ur Kernphysik, Johannes
Gutenberg-Universit\"at, D-55099 Mainz }
\begin{document}

\def\dsdt{$\frac{d\sigma}{dt}$}
\def\beqn{\begin{eqnarray}}
\def\eeqn{\end{eqnarray}}
\def\barr{\begin{array}}
\def\earr{\end{array}}
\def\btab{\begin{tabular}}
\def\etab{\end{tabular}}
\def\bite{\begin{itemize}}
\def\eite{\end{itemize}}
\def\bcen{\begin{center}}
\def\ecen{\end{center}}

\def\eq{\begin{equation}}
\def\ee{\end{equation}}
\def\eqa{\begin{eqnarray}}
\def\eea{\end{eqnarray}}

\def\sl#1{\slash{\hspace{-0.2 truecm}#1}}

\abstract { The relativistic amplitudes of pion photoproduction are evaluated by
dispersion relations at $t=const$. The imaginary parts of the amplitudes are taken from
the MAID model covering the absorption spectrum up to center-of-mass energies
$W=2.2$~GeV. For sub-threshold kinematics the amplitudes are expanded in powers of the
two independent variables $\nu$ and $t$ related to energy and momentum transfer.
Subtracting the loop corrections from this power series allows one to determine the
counter terms of covariant baryon chiral perturbation theory. The proposed continuation
of the amplitudes into the unphysical region provides a unique framework to derive the
low-energy constants to any given order as well as an estimate of the higher order terms
by global properties of the absorption spectrum.
} 
\maketitle
%

\section{Introduction}

In a recent contribution we studied the Fubini-Furlan-Rossetti (FFR) sum rule,
which relates the anomalous magnetic moment to single-pion photoproduction on
the nucleon~\cite{Pas05}. This sum rule was derived on the basis of current
algebra and PCAC in the chiral limit of massless pions~\cite{Fub66}. It
requires a continuation of the production amplitude to the threshold kinematics
of massless pions, which is of course outside of the physical region. We
evaluated the FFR sum rule by dispersion relations (DRs) at $t=const$, using
the imaginary parts of the MAID model~\cite{MAID} as input for the dispersion
integrals. As was to be expected, we obtained large corrections to the sum rule
in the physical threshold region due to the finite pion mass. However, the FFR
sum rule was found to be closely obeyed by the production amplitude in the
unphysical region close to the threshold for massless pions.

Our further discussion is facilitated by a look at the Mandelstam plane shown in
fig.~\ref{mandelstam_pm} for the isovector photon. The two independent kinematical
variables are chosen to be $\nu$ and $t$, related to energy and momentum transfer,
respectively (see section~\ref{inv_amp} for a detailed discussion). The physical region
lies between the solid lines labeled $\theta=0$ (forward scattering) and
$\theta=180^{\circ}$ (backward scattering), which meet at the production threshold
described by $\nu=\nu_{\rm{thr}}\approx M_\pi$ and $t=t_{\rm{thr}}\approx -M_\pi^2$,
where $M_\pi$ is the pion mass. In the case of a massless pion the threshold moves to the
origin of the Mandelstam plane, and it is the pion production amplitude at this point
that is related to the anomalous magnetic moment by the FFR sum rule.

The triangle near the origin is the region where no production can take place.
It is bounded by the dotted lines $s=(M_N+M_\pi)^2$ and, for the isovector
photon, $t=9M_\pi^2$. In the former case, the variable $s$ indicates the square
of the center-of-mass (c.m.) energy in the ``s-channel'' $\gamma+N\to \pi+N$,
which has to be large enough to produce a pion and a nucleon with mass $M_N$.
The other boundary refers to the crossed or ``t-channel'' $\gamma+\pi\to
N\bar{N}$ whose lowest inelasticity is due to the reaction $\gamma+\pi\to 3\pi$
requiring a c.m. energy of at least three pion masses.

Whereas the production amplitudes in the described triangle can be defined as
real functions, they become complex outside of the triangles due to
rescattering or competing reactions, and therefore the boundaries are lines of
singularities for these functions. Further singularities appear for
intermediate one-nucleon and one-pion states in the form of pole terms, e.g.,
$s=M_N^2$ and $t=M_\pi^2$. Since these pole terms are well known we subtract
them from the full amplitudes and refer to the remaining terms as the
dispersive amplitudes. These amplitudes can then be expanded in a real power
series in $\nu$ and $t$, with a convergence radius given by the onset of
pion-nucleon rescattering or the opening of two- and three-pion t-channel
reactions for the isoscalar and isovector photon, respectively.

It is the aim of our present paper to derive the power series for the pion
photoproduction amplitudes with MAID05 as an input for the imaginary parts, and
thus to provide a framework that determines the low-energy constants (LECs) of
chiral effective field theories by global properties of the nucleon's
excitation spectrum.

The general structure of the relativistic amplitudes was already discussed in the early
1990's in the context of a then existing puzzle for neutral pion photoproduction at
threshold~\cite{Bec90}. By explicitly calculating the pion loop diagrams, Bernard et
al.~\cite{Ber91} showed that the Taylor coefficients of the expansion in $\nu$ and $t$
were divergent in the limit of massless pions, which invalidated the proofs for a
low-energy theorem that was based on non-singular coefficients in the chiral limit. Since
there appeared to be complications in the power counting for the relativistic theory, the
problem was then reformulated in heavy baryon chiral perturbation theory (HBChPT), which
allowed for a strict correspondence between the loop expansion and the expansion in small
external momenta and quark (or pion) masses. Being based on a $1/M_N$ expansion, the
framework of HBChPT leads however to shifts of the pole positions, with the result of a
spurious behavior of the amplitudes in the unphysical region near the origin of the
Mandelstam plane~\cite{Pas05}. These shortcomings of HBChPT have been noted often before,
and the newly developed manifestly Lorentz-invariant renormalization schemes~\cite{Bec99}
now provide a covariant treatment as well as a consistent ordering scheme. In particular
Bernard et al.~\cite{Ber05} have recently analyzed the FFR sum rule in the framework of
infrared regularization of BChPT and obtained both a good agreement with the threshold
data and the expected smooth $\nu$ dependence of the amplitudes in the (unphysical)
sub-threshold region.

We proceed in section~\ref{inv_amp} by defining the kinematical variables and recalling
the invariant amplitudes and their dependence on the multipoles. In the following
section~\ref{sec:dr} we discuss the convergence of DRs at $t=const$ and show that the
extrapolation of the amplitudes into the region near $\nu=t=0$ should not be a problem.
In section 4 we discuss the special role of t-channel vector meson exchange and the
high-energy contribution to the dispersion integral. Our results are then presented in
section~\ref{sec:results}, together with a comparison to the data and to covariant BChPT.
In section~\ref{sec:sum} we close with a short summary and an outlook.

\section{Kinematics and invariant amplitudes \label{inv_amp}}

Let us first define the kinematics of pion photoproduction on a nucleon, the
reaction
\[
\gamma (k) + N(p_i)\to\pi(q) + N'(p_f)\,,
\]
where the variables in brackets denote the four-momenta of the participating
particles. The familiar Mandelstam variables are
\begin{eqnarray}
s=(p_i+k)^2,\quad t=(q-k)^2,\quad u=(p_i-q)^2,
\end{eqnarray}
and
\begin{eqnarray}
\nu=(s-u)/4M_N
\end{eqnarray}
is the crossing symmetrical variable. This variable is related to the photon
lab energy $E_\gamma^{lab}$ by
\begin{eqnarray}
\nu=E_\gamma^{lab} + \frac{t-M_\pi^2}{4M_N}\,.
\end{eqnarray}
The physical s-channel region is shown in fig.~\ref{mandelstam_pm}. Its upper and lower
boundaries are given by the scattering angles $\theta=0$ and $\theta=180^{\circ}$,
respectively. The pion and nucleon poles lie in the unphysical region on the straight
lines $t=M_\pi^2$ (pion pole), $\nu_s=\nu_B$ (s-channel nucleon pole), and $\nu_u=-\nu_B$
(u-channel nucleon pole), where
\begin{eqnarray}
\nu_B =\frac{t-M_\pi^2}{4M_N}\,.
\end{eqnarray}
We note that all the involved particles are on their mass shell at the point
$(\nu=0,\,t=M_\pi^2)$.

The threshold for pion photoproduction lies at
\begin{eqnarray}
\nu_{{\rm thr}}
&=& \frac{M_{\pi} (2M_N+M_\pi)^2}{4M_N (M_N+M_\pi)}\,,\nonumber\\
t_{{\rm thr}} &=& -\frac{M_{\pi}^2 M_N}{M_N+M_\pi}\,.
\label{threshold_PP}
\end{eqnarray}

In the pion-nucleon c.m. system, we have
\begin{equation}
\begin{array}{rllrll}
p_i^\mu & = & (E_i, -\vec k)\,, & p_f^\mu & = & (E_f, -\vec q)\,,\nonumber\\
k^\mu & = & (|\vec k|, \vec k)\,, &  q^\mu & = & (\omega, \vec q)\,,
\end{array}
\end{equation}
where
\begin{eqnarray}
\label{eq:kinematics} k&=&|\vec k|=\frac{s-M_N^2}{2\sqrt{s}}\,,\quad
\omega=\frac{s+M_\pi^2-M_N^2}{2\sqrt{s}}\,,\nonumber\\
q & = & |\vec q|=\left[\left(\frac{s+M_\pi^2-M_N^2}{2\sqrt{s}}\right)^2
-M_\pi^2\right]^{1/2}\nonumber \\
& = &\left[\left(\frac{s-M_\pi^2+M_N^2}{2\sqrt{s}}\right)^2
-M_N^2\right]^{1/2}\,,\nonumber\\
E_i & = & W-k=\frac{s+M_N^2}{2\sqrt{s}}\,, \nonumber \\
E_f & = &W-\omega=\frac{s+M_N^2-M_\pi^2}{2\sqrt{s}}\,,
\end{eqnarray}
with $W=\sqrt{s}$ the c.m. energy.

The transition current operator for pion photoproduction can be expressed in terms of 4
invariant amplitudes $A_i$~\cite{Che57,Han98},
\begin{eqnarray}
\label{eq:inv_ampl}
J^\mu = \sum_i A_i(\nu,t)\, M^\mu_i,
\end{eqnarray}
with the four-vectors $M^\mu_i$ given by
\begin{eqnarray}
M^\mu_1&=&
-\frac{1}{2}i\gamma_5\left(\gamma^\mu\sl{k}-\sl{k}\gamma^\mu\right)\, ,
\nonumber\\
M^\mu_2&=&2i\gamma_5\left(P^\mu\, k\cdot q-
q^\mu\,k\cdot P\right)\, ,\nonumber\\
M^\mu_3&=&-i\gamma_5\left(\gamma^\mu\, k\cdot q
-\sl{k}q^\mu\right)\, ,\nonumber\\\
M^\mu_4&=&-2i\gamma_5\left(\gamma^\mu\, k\cdot P -\sl{k}P^\mu\right)-2M_N \,
M^\mu_1\, ,
\label{eq:tensor}
\end{eqnarray}
where $P^\mu=(p_i^\mu+p_f^\mu)/2$ and the gamma matrices are defined as in
Ref.~\cite{Bjo65}.

The invariant amplitudes can be further decomposed into three isospin channels
$A_i^I(I=+,0,-)$,
\begin{eqnarray}
A_i^a=A_i^{(-)}i\epsilon^{a3b}\tau^b+A_i^{(0)}\tau^a+A_i^{(+)}\delta_{a3},
\end{eqnarray}
where $\tau^a$ are the Pauli matrices in isospace. The physical photoproduction
amplitudes are then obtained from the following linear combinations:
\begin{eqnarray}
A_i(\gamma p\rightarrow n\pi^+)&=&\sqrt{2}(A_i^{(-)}+A_i^{(0)}),\nonumber\\
A_i(\gamma p\rightarrow p\pi^0)&=&A_i^{(+)}+A_i^{(0)},\nonumber\\
A_i(\gamma n\rightarrow p\pi^-)&=&-\sqrt{2}(A_i^{(-)}-A_i^{(0)}),\nonumber\\
A_i(\gamma n\rightarrow n\pi^0)&=&A_i^{(+)}-A_i^{(0)}.
\label{linear_comb}
\end{eqnarray}
The FFR sum rule is derived from the neutral pion photoproduction amplitude in the limit
of $q^\mu \rightarrow 0$. As we note from eq.~(\ref{eq:tensor}), the four-vectors
$M^\mu_2,$ $M^\mu_3,$ and $M^\mu_4$ vanish, and only the four-vector $M^\mu_1$ survives
in that limit.

Corresponding to their behavior under crossing, the amplitudes
$A_{1,2,4}^{(+,0)}$ and $A_{3}^{(-)}$ are even functions of $\nu$ and satisfy a
DR of the type
\begin{eqnarray}
\label{eq:dr1}
&&{\rm Re}\,A^{I}_i(\nu,t)= \nonumber \\
&&A_i^{I\,,\rm{pole}}(\nu,t) +\frac{2}{\pi}{\cal
P}\int_{\nu_{thr}}^{\infty}{\rm d}\nu' \frac{\nu'\,{\rm
Im}\,A_i^I(\nu',t)}{\nu'^2-\nu^2}\,,
\end{eqnarray}
whereas the amplitudes $A_3^{(+,0)}$ and $A_{1,2,4}^{(-)}$ are odd and
therefore fulfil the relation
\begin{eqnarray}
\label{eq:dr2}
&&{\rm Re}\,A^I_i(\nu,t)= \nonumber \\
&&A_i^{I\,,\rm{pole}}(\nu,t) +\frac{2\nu}{\pi}{\cal
P}\int_{\nu_{thr}}^{\infty}{\rm d}\nu' \frac{{\rm
Im}\,A_i^I(\nu',t)}{\nu'^2-\nu^2}\,.
\end{eqnarray}
The nucleon and pion pole contributions are given by
\begin{eqnarray}
A_1^{I,pole} & = & \ \ \ \frac{eg_{\pi N}}{2}
\left(\frac{1}{s-M_N^2}+\frac{\epsilon^I}{u-M_N^2}\right)\,,\nonumber \\
A_2^{I,pole} & = & -\frac{eg_{\pi N}}{t-m^2_\pi}
\left(\frac{1}{s-M_N^2}+\frac{\epsilon^I}{u-M_N^2}\right)\,,\nonumber \\
A_3^{I,pole} & = & -\frac{eg_{\pi N}}{2M_N}\frac{\kappa^{I}}{2}
\left(\frac{1}{s-M_N^2}-\frac{\epsilon^I}{u-M_N^2}\right)\,,\nonumber \\
A_4^{I,pole} & = & -\frac{eg_{\pi N}}{2M_N}\frac{\kappa^{I}}{2}
\left(\frac{1}{s-M_N^2}+\frac{\epsilon^I}{u-M_N^2}\right)\,,
\label{eq:a1-4pole}
\end{eqnarray}
with $\epsilon^+=\epsilon^0=-\epsilon^-=1$, $\kappa^{(+,-)}= \kappa_p-\kappa_n$, and
$\kappa^{(0)}=\kappa_p+\kappa_n$, where $\kappa_p$ and $\kappa_n$ are the anomalous
magnetic moments of the proton and the neutron, respectively. Additional pole
contributions from the t-channel vector meson exchange are discussed in
section~\ref{sec:vec_mes}.

The covariant amplitudes $A_i$ can be expressed by the CGLN amplitudes
$\mathcal{F}_i$~\cite{Che57,Han98} as follows:
\begin{eqnarray}
\label{eq:F1}
\lefteqn{ A_1 =  {\mathcal N}\  \bigg\{ \frac{W+M_N}{W-M_N}\,\mathcal{F}_1 -
(E_f+M_N)\,\frac{\mathcal{F}_2}{q} }
\vspace{0.3cm} \nonumber \\
&& + \frac{M_N(t-M_\pi^2)}{(W-M_N)^2}\,\frac{\mathcal{F}_3}{q} +
\frac{M_N(E_f+M_N)\,(t-M_\pi^2)}{W^2-M_N^2}\,\frac{\mathcal{F}_4}{q^2}
\bigg\}\, ,\nonumber \\
&&  \\
\label{eq:F2}
\lefteqn{ A_2 = \frac{{\mathcal N}}{W-M_N} \left\{ \frac{{\mathcal F}_3}{q} -
(E_f+M_N)\,\frac{{\mathcal F}_4}{q^2} \right\}\,,}
\vspace{0.3cm} \nonumber \\
&& \\
\label{eq:F3}
\lefteqn{ A_3 =  \frac{{\mathcal N}}{W-M_N}\  \bigg\{ \mathcal{F}_1 +
(E_f+M_N)\ \frac{\mathcal{F}_2}{q}  }
\vspace{0.3cm} \nonumber \\
&&  +\left( W+M_N + \frac{t-M_\pi^2}{2(W-M_N)} \right)\ \frac{\mathcal{F}_3}{q}
\nonumber \\
&&  +\left( W-M_N + \frac{t-M_\pi^2}{2(W+M_N)}
\right)\, (E_f+M_N)\,\frac{\mathcal{F}_4}{q^2} \bigg\}\,, \nonumber \\
&& \\
\label{eq:F4}
\lefteqn{ A_4 =  \frac{{\mathcal N}}{W-M_N} \bigg\{ \mathcal{F}_1 +
(E_f+M_N)\,\frac{\mathcal{F}_2}{q}  }
\vspace{0.3cm} \nonumber \\
&& +\frac{t-M_\pi^2}{2(W-M_N)}\,\frac{\mathcal{F}_3}{q} +
\frac{t-M_\pi^2}{2(W+M_N)}\, (E_f+M_N)\,\frac{\mathcal{F}_4}{q^2} \bigg\}
\nonumber \,, \\
\end{eqnarray}
where ${\mathcal N} = 4\pi/\sqrt{(E_i+M_N)\,(E_f+M_N)}$, $q=|\vec{q}|$ and all variables
are expressed in the c.m. frame. Below the $\Delta(1232)$ resonance, we may limit
ourselves to the S-wave multipole $E_{0+}$ and to the three P-wave multipoles $E_{1+},\
M_{1+}$, and $M_{1-}$. In this approximation, the CGLN amplitudes take the form
\begin{equation}
\begin{array}{rllrll}
\mathcal{F}_1 & \to & E_{0^+} + 3(M_{1^+} + E_{1^+})\cos\theta\,,
\\ \\
\mathcal{F}_2 & \to & 2M_{1^+}+M_{1^-}\ ,
\\ \\
\mathcal{F}_3 & \to & 3(E_{1^+}-M_{1^+})\ , \quad & \mathcal{F}_4& \to &0 \ ,
\end{array}
\end{equation}
where $\theta$ is the c.m. scattering angle, which is related to the Mandelstam
variables by
\begin{equation}
\label{theta}
\cos\theta=\frac{(s-M_N^2)^2-M_\pi^2(s+M_N^2)+2\,s\,t}{2\,q\,\sqrt{s} \,
(s-M_N^2)} \ .
\end{equation}
The P-wave contributions are often expressed by the three combinations
\begin{eqnarray}
{P}_1 & = & 3E_{1+}+M_{1+} - M_{1-}\ ,  \nonumber \\
{P}_2 & = & 3E_{1+}-M_{1+} + M_{1-}\ ,\nonumber \\
{P}_3 & = & 2M_{1+} + M_{1-}\ .
\end{eqnarray}
With these definitions the multipole expansion of eqs.
(\ref{eq:F1})~-\,(\ref{eq:F4}) can be cast into the form
\begin{eqnarray}
\label{eq:A1}
 A_1  & = & {\mathcal N} \, \frac{W+M_N}{W-M_N}\
 \bigg\{ E_{0+} +\left( \omega + \frac{W(t-M^2_\pi)}{W^2-M_N^2}\right
)\,\bar{P}_1 \nonumber \\
&&  +   \frac{M_N(t-M_\pi^2)}{W^2-M_N^2}\,\bar{P}_2 +
\frac{t}{W+M_N}\,\bar{P}_3 + \ldots \bigg\}\ , \nonumber \\
&& \\
\label{eq:A2}
A_2 & = &  \frac{{\mathcal N}}{W-M_N}\  \{\bar{P}_2 - \bar{P}_3 + \ldots \}
\,,\nonumber \\
&& \\
\label{eq:A3}
A_3 & = & \frac{{\mathcal N}}{W-M_N}\  \bigg\{ E_{0+} + \left( \omega +
\frac{W(t-M^2_\pi)}{W^2-M_N^2}\right )\,\bar{P}_1 \nonumber \\
&& + \left ( W+M_N + \frac{t-M_\pi^2}{2(W-M_N)} \right )\,\bar{P}_2
 \nonumber \\
&&   + \frac{t-M_\pi^2}{2(W+M_N)}\,\bar{P}_3 + \ldots \bigg\}\,,\nonumber \\
&& \\
\label{eq:A4}
A_4 & = & \frac{{\mathcal N}}{W-M_N}\ \bigg\{ E_{0+}  \nonumber \\
&& +\left( \omega + \frac{W(t-M^2_\pi)}{W^2-M_N^2}\right )\,\bar{P}_1 +
\frac{(t-M_\pi^2)}{2(W-M_N)}\,\bar{P}_2  \nonumber \\
&&  + \left ( W+M_N +
\frac{t-M_\pi^2}{2(W+M_N)} \right )\, \bar{P}_3 + \ldots \bigg \} \,,\nonumber \\
\end{eqnarray}
with $\bar{P}_i = P_i/q$ and the ellipses denoting the higher partial waves.

The kinematical factors simplify at threshold, and eqs.~(\ref{eq:A1}) -~(\ref{eq:A4})
take the exact form
\begin{eqnarray}
\label{eq:a1_thr}
\lefteqn {A_1(\nu_{\rm{thr}}, t_{\rm{thr}})  = {\mathcal N}_{\rm{thr}}}
\nonumber \\
&&\times\bigg\{ E_{0+}-\frac{M_\pi}{1+\mu}\,\bar{P}_2 -
\frac{\mu\,M_\pi}{(1+\mu)(2+\mu)}\, \left (\bar{P}_3 +6M_N\,\bar{D} \right
)\bigg\}\,, \nonumber \\
&& \\
\label{eq:a2_thr}
\lefteqn{ A_2(\nu_{\rm{thr}}, t_{\rm{thr}}) = \frac{{\mathcal N}_{\rm{thr}}}
{M_N(2+\mu)}\  \bigg \{ \bar{P}_2 - \bar{P}_3 - 6M_N \bar{D}
\bigg \}\,,} \nonumber \\
&& \\
\label{eq:a3_thr}
\lefteqn {A_3(\nu_{\rm{thr}}, t_{\rm{thr}}) = \frac{{\mathcal N}_{\rm{thr}}}
{M_N(2+\mu)}\  \bigg\{ E_{0+} + \frac{(2+\mu)^2M_N}{2(1+\mu)}\bar{P}_2 }
\vspace{0.3cm} \nonumber \\
&&  - \frac{\mu\,M_\pi}{2(1+\mu)}\,\bar{P}_3 +
\frac{3(2+\mu)M_NM_\pi}{1+\mu}\,\bar{D}\bigg \}\,, \nonumber \\
&& \\
\label{eq:a4_thr}
\lefteqn {A_4(\nu_{\rm{thr}}, t_{\rm{thr}}) = \frac{{\mathcal N}_{\rm{thr}}}
{M_N(2+\mu)}\ \bigg\{ E_{0+} - \frac{(2+\mu)M_\pi}{2(1+\mu)}\bar{P}_2 }
\vspace{0.3cm} \nonumber \\
&&  + \frac{(4+6\mu+\mu^2)\,M_N}{2(1+\mu)}\,\bar{P}_3 -
\frac{3M^2_\pi}{1+\mu}\,\bar{D}\bigg \}\,, \nonumber \\
\end{eqnarray}
where ${\mathcal N}_{\rm{thr}}=4\pi\sqrt{1+\mu}/M_\pi$, $\mu=M_\pi/M_N$,
$\bar{D}=(M_{2+}-E_{2+}-M_{2-}-E_{2-})/q^2$, and all the multipoles have to be
evaluated at $q=0$. In particular we note that the amplitude $\bar{P}_1$ does
not appear in eqs.~(\ref{eq:a1_thr})-(\ref{eq:a4_thr}), because its kinematical
prefactor vanishes at threshold.

Having determined the invariant amplitudes $A_i$, we can combine these results
to construct the CGLN amplitudes ${\mathcal F}_i$ as follows:
\begin{eqnarray}
\label{eq:CGLN1}
\lefteqn { {\mathcal F}_1  =  \sqrt{(E_i+M_N)(E_f+M_N)}\,\frac{W-M_N}{8\pi\,W}
} \vspace{0.3cm} \nonumber \\
&& \times\bigg\{ A_1 + (W-M_N)\,A_4 - \frac{t-M_\pi^2}{2(W-M_N)}\,(A_3-A_4)
\bigg\}\,, \nonumber \\
&& \\
\label{eq:CGLN2}
\lefteqn { {\mathcal F}_2  =
\sqrt{\frac{E_i-M_N}{E_f+M_N}}\,\frac{W+M_N}{8\pi\,W}\,q
} \vspace{0.3cm} \nonumber \\
&& \times\bigg\{ -A_1 + (W+M_N)\,A_4 - \frac{t-M_\pi^2}{2(W+M_N)}\,(A_3-A_4)
\bigg\}\,, \nonumber \\
&& \\
\label{eq:CGLN3}
\lefteqn { {\mathcal F}_3  =
\sqrt{(E_i-M_N)(E_f+M_N)}\,\frac{W+M_N}{8\pi\,W}\,q }
\vspace{0.3cm} \nonumber \\
&& \times\bigg\{ (W-M_N)\,A_2 + A_3-A_4 \bigg \}\,, \nonumber \\
&& \\
\label{eq:CGLN4}
\lefteqn { {\mathcal F}_4  =
\sqrt{\frac{E_i+M_N}{E_f+M_N}}\,\frac{W-M_N}{8\pi\,W}\,q^2 }
\vspace{0.3cm} \nonumber \\
&& \times\bigg \{ - (W+M_N)\,A_2 +
A_3 - A_4 \bigg \}\,.  \nonumber \\
\end{eqnarray}

\section{Convergence of dispersion relations at t=const \label{sec:dr}}

We recall that the integration path for DRs at $t=const$ is fully contained in the
physical region only in the special case of $t=t_{\rm{thr}}$, which path passes through
the threshold of pion photoproduction ($\nu=\nu_{\rm{thr}},\ t=t_{\rm{thr}}$). In all
other cases the dispersion integrals run through an unphysical part of the Mandelstam
plane between $s=(M_N+M_\pi)^2$ and the physical threshold value for $s$, which
corresponds to forward scattering ($\theta=0$) for $t>t_{\rm{thr}}$ and backward
scattering ($\theta=180^o$) for $t<t_{\rm{thr}}$. The only successful procedure to
construct ${\rm Im}\,A_i(s,t)$ in the unphysical region has been the multipole expansion
of the CGLN amplitudes and the subsequent insertion of ${\rm Im}\,\mathcal{F}_i(s,t)$
into eqs.~(\ref{eq:F1})-(\ref{eq:F4}). What about the convergence of this expansion?
According to Mandelstam~\cite{Man58} the scattering amplitudes can be represented by the
pole terms and a double integral over the spectral regions $A_{su}$, $A_{ts}$, and
$A_{st}$ shown in fig.~\ref{mandelstam_pm}. The boundaries of these regions result from
the lowest mass intermediate states possible in any pair of the variables $s,\ t,$ and
$u$. We hasten to add that ``possible'' refers to solutions with real values of $s,\ t,$
and $u$ and all particles on their mass shell, however there is no overlap between the
double spectral regions and the physical regions. Therefore the Mandelstam representation
has never had any direct consequences for the data analysis. This representation has
however important consequences because it reflects maximum analyticity, in the sense that
the only singularities are given by the poles due to one-particle intermediate states and
the cuts due to the onset of particle production channels. In particular, one-dimensional
DRs such as DRs at $t=const$ can be straightforwardly derived from the Mandelstam
representation~\cite{Fra60}. After subtraction of the pole terms from the full amplitudes
$A_i$, the range of convergence is determined by the nearest singularity in the
Mandelstam plane, which arrives when the line $t=const$ becomes tangent to the double
spectral region~\cite{Bal61}. In the case of the multipole expansion, the convergence is
based on the following mathematical lemma: If a function $f(z=x+iy)$ is analytic inside
and on an ellipse $C$ whose foci are at the points ($x=\cos\theta=\pm1,\,y=0$), it can be
expanded in a Legendre series for all points in the interior of the ellipse
$C$~\cite{Whi58}.

The relevant relations for $t$ may be obtained from eq.~(\ref{theta}) and the
kinematics of eq.~(\ref{eq:kinematics}),
\begin{equation}
t(s)=M_\pi^2-2\omega(s)k(s) + 2q(s)k(s)\cos\theta\,.
\end{equation}
The center of the ellipse lies on the line $\theta=90^{\circ}$,
\begin{eqnarray}
\label{t0}
t_0(s) & = & M_\pi^2-2\omega(s)k(s)\,,
\end{eqnarray}
and the foci correspond to forward and backward scattering,
\begin{eqnarray}
t_\pm(s) & = & t_0(s) \pm2q(s)k(s)\,.
\end{eqnarray}
The ellipse of convergence can now be stretched until its upper or lower
tangent $t=const$ touches the nearest double spectral region of the Mandelstam
representation. For the isovector amplitudes the boundary of the double
spectral region $A_{st}^{(\pm)}$ is given by
\begin{eqnarray}
\label{t_upper_s}
t_{\rm{upper}}(s) =
9M_\pi^2+\frac{8M_\pi^4\,(3s-M_N^2+M_\pi^2)}{[s-(M_N+M_\pi)^2]\,[s-(M_N-M_\pi)^2]}\,,\nonumber
\\
\end{eqnarray}
with the asymptote $t_{\rm{upper}}(s\to\infty)=9M_\pi^2$ as the upper limit of
convergence of DRs at $t=const$. The lower limit is given by the reflection of
eq.~(\ref{t_upper_s}) at the center of the ellipse, eq.~(\ref{t0}). This leads
to
\begin{eqnarray}
\label{t_lower_s}
t_{\rm{lower}}(s) = 2t_0(s) - t_{\rm{upper}}(s)\,.
\end{eqnarray}
The maximum of this function lies at $s\approx1.32$~GeV$^2$ and takes the values
$t_{\rm{min}}=-0.376$~GeV$^2$ and $-0.352$~GeV$^2$ for charged and neutral pions,
respectively.

The corresponding boundary for the isoscalar amplitudes $A_{st}^{(0)}$ is defined by the
lower value of the two curves
\begin{eqnarray}
\label{t_upper_s i} t_{\rm{upper}}^{(i)}(s) = Q_i(s) + \sqrt{Q_i^2(s)+R_i(s)}\,,\quad
i=1\
\rm{and}\ 2\,, \nonumber \\
\end{eqnarray}
with
\begin{eqnarray}
\label{QQ1} Q_1(s) & = & 2M_\pi^2 +
\frac{(9s+31M_N^2-28M_\pi^2)M_\pi^4}{[s-(M_N+2M_\pi)^2][s-(M_N-2M_\pi)^2]}\,,\nonumber\\
&&\\
\label{QQ2} Q_2(s) & = & 8M_\pi^2 +
\frac{4(9s+M_N^2-M_\pi^2)M_\pi^4}{[s-(M_N+M_\pi)^2][s-(M_N-M_\pi)^2]}\,,\nonumber
\\
&& \\
\label{RR1} R_1(s) & = &
\frac{(4M_N^2-M_\pi^2)M_\pi^6}{[s-(M_N+2M_\pi)^2][s-(M_N-2M_\pi)^2]}\,,\nonumber
\\
&& \\
\label{RR2} R_2(s) & = &
\frac{16(M_N^2-M_\pi^2)M_\pi^6}{[s-(M_N+M_\pi)^2][s-(M_N-M_\pi)^2]}\,.\nonumber \\
\end{eqnarray}
The first of these curves defines the upper limit of convergence by the
asymptote $t_{\rm{upper}}^{(1)}(s\to\infty)=4M_\pi^2$. The lower limit is again
obtained by reflection according to eq.~(\ref{t_lower_s}), and its maximum at
$s\approx1.65$~GeV$^2$ yields the lower limit of convergence at
$t_{\rm{min}}=-0.525$~GeV$^2$. Altogether then we find that DRs should converge
in the following strip of the $\nu-t$ plane:
\begin{eqnarray}
\label{nu_t_plane}
A_i^{(\pm)} & : & -0.376~{\rm GeV}^2 < t< 0.175~{\rm GeV}^2  \\
A_i^{(0)} & : & -0.525~{\rm GeV}^2 < t< 0.073~{\rm GeV}^2 \,.\nonumber
\end{eqnarray}
For completeness we recall two other critical values below these limits. With
decreasing $t$ the s- and u-channel cuts approach and touch at $\nu=0$,
$t=-2M_\pi(2M_N+M_\pi)\approx-0.564$~GeV$^2$. Up to this point, and with some
care also in the region below, DRs should still be valid, except that the
Legendre expansion is no longer convergent. The final break-down of DRs occurs
if $t=const$ is tangent to the double spectral region $A_{su}^{(\pm,0)}$ given
by the relation
\begin{eqnarray}
\label{su_spectralr}
&& [s-(M_N+M_\pi)^2] [s-(M_N-M_\pi)^2] [u-(M_N+M_\pi)^2]  \nonumber \\
&& \times[u-(M_N-M_\pi)^2]  - (4M_N^2-M_\pi^2)M_\pi^2 \nonumber
\\
&& \times[2su-2(M_N^2-M_\pi^2) (s+u) + 2M_N^4-M_\pi^4] =0\,.
\end{eqnarray}
This happens at $\nu=0$ and $t=-1.10$~GeV$^2$ and $-1.06$~GeV$^2$ for charged
and neutral pions, respectively.

For completeness we also mention the work of Oehme and Taylor~\cite{Oeh59}. On the basis
of relativistic quantum field theory, especially causality and the spectral conditions,
these authors have rigorously proved that a polynomial expansion in $\cos\theta$
converges for $-12M_\pi^2<t<0$, which yields a considerably smaller range than in the
case of the Legendre expansion of Ref.~\cite{Bal61}.

\section{Vector mesons and Regge tails\label{sec:vec_mes}}

The t-channel exchange of vector mesons plays an important role in neutral pion
photoproduction. The vector meson contributions to the invariant amplitudes
take the form:
\begin{eqnarray}
A_1^V (t) & = & \frac{e\lambda_V\,g_V^{(T)}}{2M_NM_\pi}\ \frac{t}{t-m_V^2}\,,
\nonumber \\
A_2^V (t) & = & -\frac{e\lambda_V\,g_V^{(T)}}{2M_NM_\pi}\ \frac{1}{t-m_V^2}\,,
\nonumber \\
A_3^V (t) & = & 0 \,, \nonumber \\
A_4^V (t) & = & - \frac{e\lambda_V\,g_V^{(V)}}{M_\pi}\ \frac{1}{t-m_V^2}\,,
\label{eq:46}
\end{eqnarray}
where $\lambda_V$ denotes the coupling of the vector meson ($V=\omega,\rho$) to the
$\gamma\pi^0$ system and $g_V^{(V,T)}$ its vector or tensor coupling to the nucleon. Due
to their isospin structures, the $\omega$ contributes to the isospin amplitude $I=+$ and
the $\rho$ to $I=0$. In the 2005 version of MAID the coupling constants for the
$\omega\,(782)$ (in bracket: $\rho\,(770)$) are: $\lambda_V=0.314\,(0.103)$,
$g_V^{(V)}=16.3\,(1.78)$, $g_V^{(T)}=-15.4\,(22.6)$. Whereas the couplings $\lambda_V$
for the $V\to\gamma\pi$ process are essentially known, the vector and tensor couplings
$g_V^{(V,T)}$ to the nucleon are less well determined. However, all analyses agree that
the $\omega$ exchange is quite essential for neutral pion photoproduction, whereas the
$\rho$ exchange is almost negligible.

In the zero-width approximation the vector meson poles at $t=m_V^2$ should play a similar
role as the nucleon poles at $\nu=\pm\nu_B(t)$ and, in the case of charged pion
production, the pion pole at $t=M_{\pi^+}^2$. All these pole terms can not be obtained by
the dispersion integrals of eqs.~(\ref{eq:dr1}) and (\ref{eq:dr2}) but have to be added
to the dispersive contributions. Of course, one has to make sure that the pole terms are
not modified by errors of numerical origin and by the derivation of the absorptive
amplitudes from the data. In order to eliminate such double-counting in the case of
charged pion electroproduction, von Gehlen~\cite{Geh69} has proposed to subtract the term
\begin{equation}
\label{gehlen}
\frac{2}{\pi}\ \frac{1}{t-M_{\pi^+}^2} \int d\nu'\ \frac{\nu' \lim_{t'\to
M_{\pi^+}^2} (t'-M_{\pi^+}^2)\, {\rm Im}\,A_i^{I}\,(\nu',t')}{\nu'^2-\nu^2}
\end{equation}
from the dispersion integral of eq.~(\ref{eq:dr1}) for the appropriate electroproduction
amplitude $(i=5,\,I=-)$. In the same spirit we have checked whether the dispersion
integral can provide a pole structure at $t=m_V^2$. The result is negative. Even though
the vector meson background plays an important role in the unitarization process of MAID,
there is no indication of a vector meson pole term similar to eq.~(\ref{gehlen}) in our
calculation. It is even more surprising that the dispersion integrals for the threshold
amplitudes change only by a few percent if we drop the vector mesons in the construction
of the absorptive amplitude, whereas the vector mesons yield 20~\% and 50~\% of the
dispersive contributions for $A_2$ and $A_4$, respectively.

In view of the problem to reproduce the vector meson poles by the dispersion integrals of
eq.~(\ref{eq:dr1}), we accept the Mandelstam hypothesis~\cite{Man58} that the amplitudes
are the sum of all pole terms plus two-dimensional integrals over the double spectral
region. The one-dimensional DRs, e.g. at $t=const$, follow from this representation, as
has been proved for pion photoproduction by Ball~\cite{Bal61}. Alternatively, we could
subtract the DRs at $\nu=0$, which introduces an unknown function $A_i^I(0,t)$. This
function is real in the region of small $t$, and in principle can be constructed from its
imaginary part~\cite{DDBPMV} by an integral along the $t$-axis (see
fig.~\ref{mandelstam_pm}). For $t>0$ the absorptive amplitude is dominated by the
reactions $\gamma + \pi \rightarrow 2 \pi$ ($I=0$) and $3 \pi$ ($I=+$) whose resonant
parts are dominated by vector meson production occurring at $t\approx
m_V^2\approx34M_\pi^2$. However, if we follow the negative $t$-axis to $t=-29 M_\pi^2$
and below we also find absorptive amplitudes, starting with $S$-wave production, followed
by $P$-wave contribution from $\Delta(1232)$ excitation, and finally effects from the
totally unknown double spectral region for $t<-58 M_\pi^2$. Unfortunately, these
contributions at $t<0$ can not be derived from the data basis, because the extrapolation
to the unphysical region by means of Legendre polynomials breaks down for $t<-19 M_\pi^2$
(see section~\ref{sec:dr}).

In particular we note that the t-channel exchange of vector mesons does not contribute to
the crossing-odd amplitudes $A_3^{(+,0)}$, which can only receive pole contributions from
the exchange of axial vector mesons.

At energies above the resonance region, the Regge model provides a convenient
description of the cross sections. It it obtained by replacing the $\rho$ and
$\omega$ propagators $(t-m_V^2)^{-1}$ by the Regge
propagators~\cite{Gui97,Azn03}
\begin{equation}
\label{regge_prop}
P_V(s,t) = \left(\frac{s}{s_0} \right)^{\alpha_V(t)-1}\frac{\pi\alpha_V
(e^{-i\pi\alpha_V(t)}-1)}{2\sin\,[\pi\alpha_V(t)]\,\Gamma\,[\alpha_V(t)]}\,,
\end{equation}
with $\alpha_V(t) = \alpha_0+\alpha_1t$ describing the Regge trajectory and $s_0$ a free
parameter. Typical values for the $\omega$ trajectory (in brackets: $\rho$ trajectory)
are $\alpha_0=0.44\ (0.55)$ and $\alpha_1=0.9$~GeV$^{-2}$ $(0.8$~GeV$^{-2})$. As a result
the Regge amplitudes have a typical $1/\sqrt{s}=1/W$ behavior for small values of $t$.

We have estimated the high-energy tails of the dispersion integrals by use of
the following models for the imaginary parts of the amplitudes:
\begin{enumerate}
\item[(I)] a $1/W$ tail fitted to MAID in the energy range of $1.8-2.5$~GeV,
\item[(II)] an energy dependence according to eq.~(\ref{regge_prop}) fitted as
above, and
\item[(III)] phenomenological parameterizations of Regge trajectories and cuts
fitted directly to older high-energy data~\cite{Dev74,Bar74}.
\end{enumerate}
While none of these descriptions is completely satisfactory, such studies
provide a reasonable estimate of the high-energy contributions.

\section{Results\label{sec:results}}

The dispersive contributions to the real part of the amplitudes $A_i╬(\nu,t)$ are
determined by the integrals of eqs.~(\ref{eq:dr1}) and (\ref{eq:dr2}). The following
figures show the respective integrands for $t=t_{\rm{thr}}$ and the two values $\nu=0$
and $\nu =\nu_{\rm{thr}}$. The dispersive amplitudes are then obtained by integration and
multiplication by the factors $2/\pi$ or $2\nu/\pi$ in front of the integrals. At this
point we remind the reader that the amplitudes $A_i$ have different dimensions due to the
traditional definitions of eqs.~(\ref{eq:inv_ampl}) and (\ref{eq:tensor}), and as a
result the integrands are given in different units. We finally note that all the
following calculations are obtained with MAID05, which now describes the amplitudes up to
$W=2.2$~GeV.

MAID is a unitary isobar model for photo- and electroproduction of pions in the resonance
region. It is constructed from a field-theoretical background with nucleon Born terms and
t-channel vector meson exchange pole terms, both unitarized in K-matrix
approximation~\cite{MAID,Maid01}. The resonance sector is modeled with s-channel nucleon
resonance excitations in Breit-Wigner form for all four-star resonances below $W=2$~GeV.
The hadronic $\rho$ and $\omega$ coupling constants of the background and the
electromagnetic resonance couplings $A_{1/2}, A_{3/2}, S_{1/2}$ that describe electric,
magnetic and Coulomb multipoles are fitted to the world data base of pion photo- and
electroproduction. Hadronic resonance parameters are taken from the Particle Data
Tables~\cite{PDG}. Latest versions of MAID (MAID03 and MAID05) describe very well the
data in the kinematical range of $W_{thr}<W<2$~GeV and
$0<Q^2<5$~GeV$^2$~\cite{Tia03,Tia04}. Over a wide energy range up to the second resonance
region MAID is generally consistent with dispersion relations at
$t=const$~\cite{DrMaid02}.

\begin{figure}[ht]
\centerline{\resizebox{0.5\textwidth}{!}{%
  \includegraphics{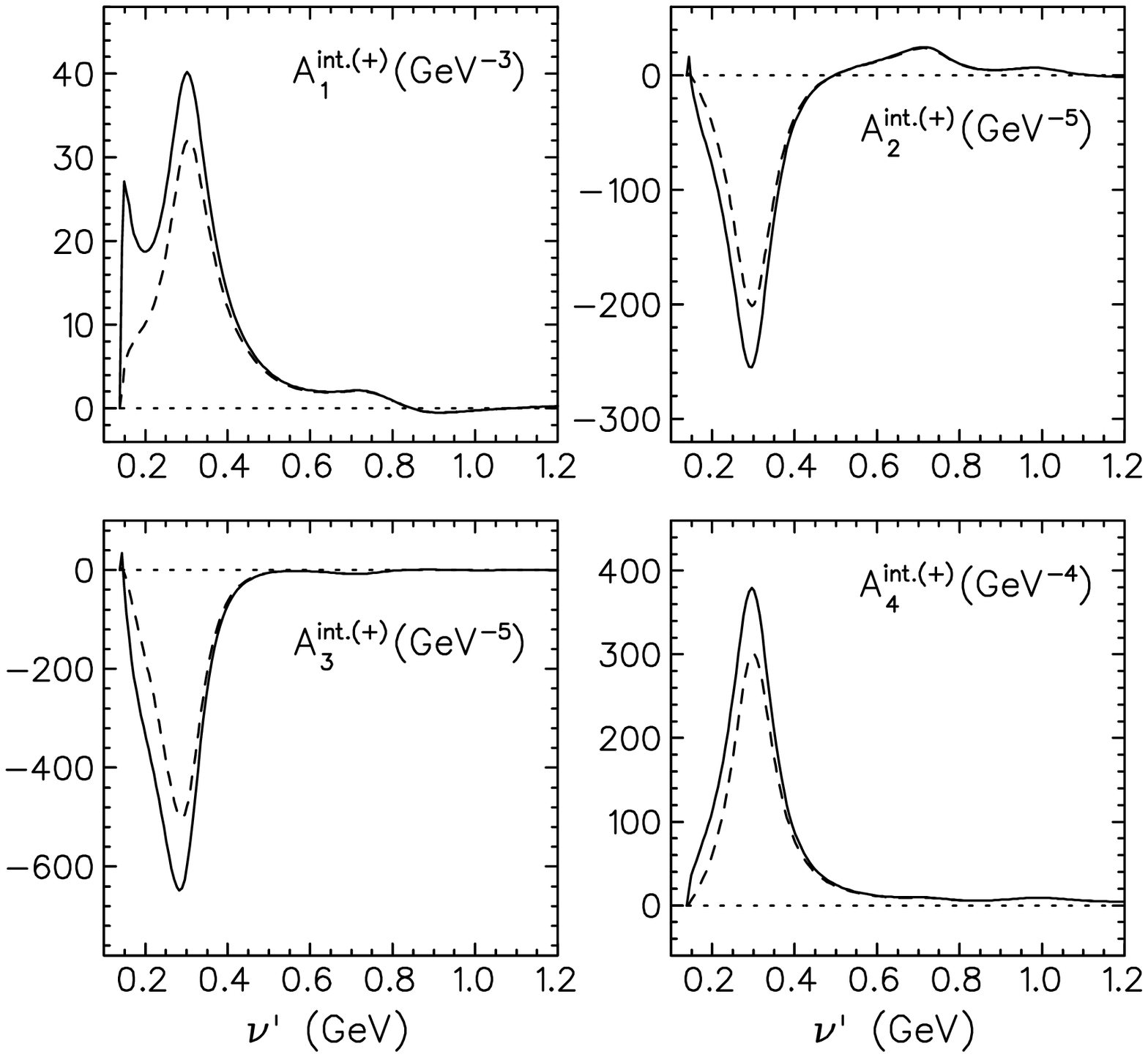}
}} \caption{ The integrands of eqs.~(\ref{eq:dr1}) and (\ref{eq:dr2}) for the amplitudes
$A_i^{(+)}$ obtained with $\nu=\nu_{\rm{thr}}$ (solid lines) and $\nu=0$ (dashed lines).}
\label{fig:integrandsplus}
\end{figure}

Figure~\ref{fig:integrandsplus} shows the integrands for the four isospin (+) amplitudes.
The dashed lines give the integrands at $\nu=0$, which are strongly dominated by the
$\Delta(1232)$ resonance with a much smaller contribution of the second resonance region
for $A_1$ and $A_2$. In addition, there appears a strong background of S-wave pion
production in the case of $A_1$. If $\nu$ approaches the onset of S-wave production (see
the full line!), the integrand increases dramatically and in principle runs into an
inverse square root singularity, which eventually leads to the cusp effect clearly seen
in $\pi^0$ production at the $\pi^+$ threshold. The cusp effect is not present for $A_2$,
because according to eq.~(\ref{eq:A1}) the S wave does not contribute to this amplitude,
and S waves are also strongly suppressed in the case of $A_3$ and $A_4$ due to
kinematical enhancement of the P-wave combinations $\bar{P}_2$ and $\bar{P}_3$,
respectively.

The integrands for the amplitudes $A_i^{(0)}$ are displayed in the following
fig.~\ref{fig:integrandszero} Since the isoscalar photon can not excite the $\Delta$
resonance, we now find a competition of contributions from the second and third resonance
regions plus a large cusp effect for $A_1^{(0)}$. Altogether these integrands are
considerably smaller than in the case of the $A_i^{(+)}$.

\begin{figure}[ht]
\centerline{\resizebox{0.5\textwidth}{!}{%
  \includegraphics{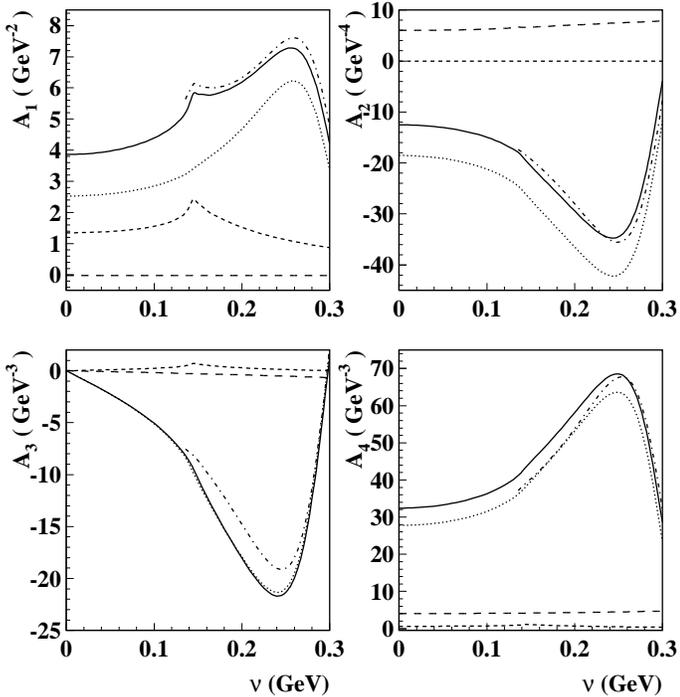}
}} \caption{ The real parts of the amplitudes $A_i^{(p\pi^0)}$ for the reaction $\gamma
p\to \pi^0p$ as function of $\nu$ and at $t=t_{\rm{thr}}$. Solid lines: Dispersive
contributions according to eqs.~(\ref{eq:dr1}) and (\ref{eq:dr2}) as obtained with the
imaginary amplitudes from MAID05 containing partial waves up to $L_{\rm{max}}=3$.
Short-dashed lines: S-wave contributions, dotted lines: P-wave contributions, long-dashed
lines: Sum of D- and F-wave contributions. The dashed-dotted lines starting at pion
threshold show the real parts taken directly from MAID05. } \label{fig:partialwaves}
\end{figure}

The presented integrands show a reasonable convergence for the higher energies. In
fig.~\ref{fig:partialwaves}  we investigate the question of convergence further by
decomposing the imaginary parts of the amplitudes into a partial wave series. The figure
shows the dispersive part of the $\gamma p\to\pi^0 p$ amplitudes as function of $\nu$ and
at $t=t_{\rm{thr}}$. While both S and P waves in the imaginary part yield large
contributions to $A_1$, the other three amplitudes are essentially determined by the P
waves, with some higher partial wave contributions in the case of $A_2$ and $A_3$.

\begin{figure}[ht]
\centerline{\resizebox{0.4\textwidth}{!}{%
  \includegraphics{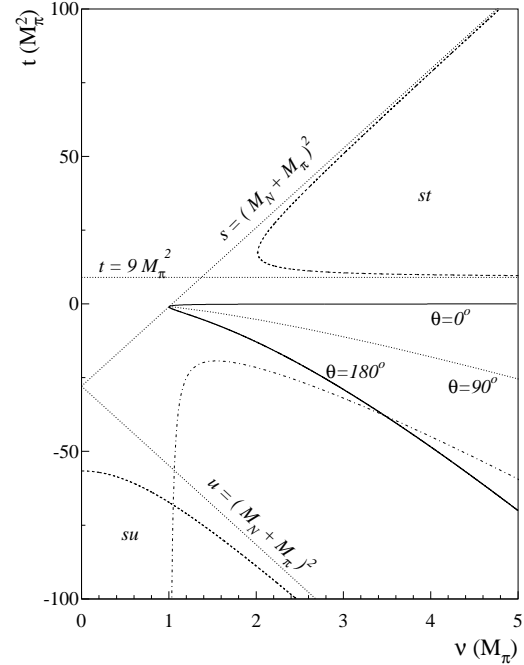}
}} \caption{ The Mandelstam plane for the isospin $I=+$ as function of $\nu$ and $t$.
Solid lines: the boundaries of the physical region between forward ($\theta=0$) and
backward ($\theta=180^o$) scattering. Dashed lines: the boundaries of the st and su
spectral regions. The dashed-dotted line is obtained by reflecting the st boundary at the
dotted line $\theta=90^o$. The other dotted lines indicate the asymptotic boundaries of
the spectral regions. The left half-plane, which is not shown here, is determined by the
crossing symmetry, $\nu\rightarrow-\nu$ or $s \leftrightarrow u$. See text for further
explanation. } \label{mandelstam_pm}
\end{figure}

\begin{figure}[ht]
\centerline{\resizebox{0.5\textwidth}{!}{%
  \includegraphics{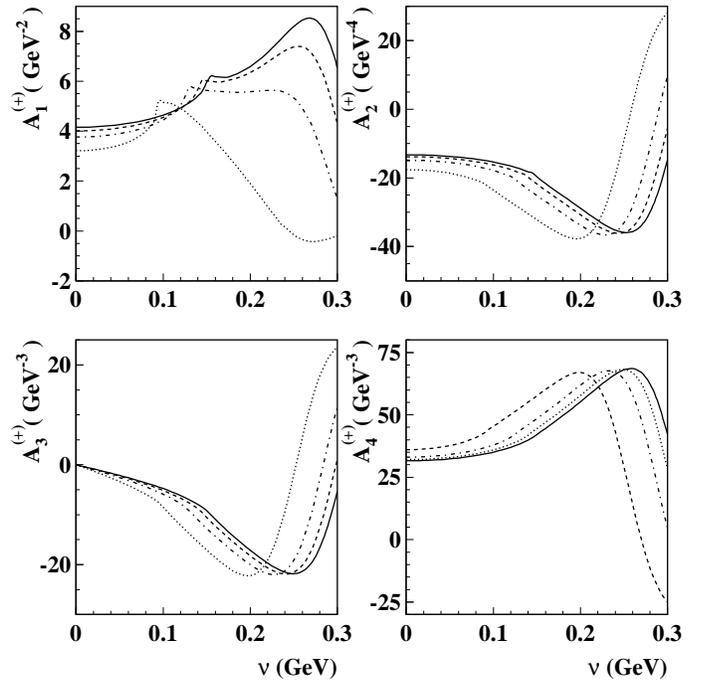}
}} \caption{ The real parts of the amplitudes $A_i^{(+)}$ as function of $\nu$. The
dispersive contributions according to eqs.~(\ref{eq:dr1}) and (\ref{eq:dr2}) as evaluated
with the imaginary amplitudes from MAID05 for the following values of $t$: $M_\pi^2$
(solid lines), $t_{\rm{thr}}$ (dashed lines), $-4M_\pi^2$ (dashed-dotted lines), and
$-10M_\pi^2$ (dotted lines).} \label{fig:tdependence}
\end{figure}

Since we want to extrapolate the amplitudes into the unphysical region at small $t$
values, we also study the dispersive parts of the isospin amplitudes as function of $\nu$
for a series of $t$ values in the range $M_\pi^2>t>-10M_\pi^2$. As shown in
fig.~\ref{fig:tdependence}, the $t$ dependence develops quite a regular pattern, with the
cusp moving to larger $\nu$ values with increasing values of $t$. At the same time the
physical threshold (see the line $\theta=180^\circ$ in fig.~\ref{mandelstam_pm}) moves to
smaller values of $\nu$ down to the minimum at $\nu=\nu_{\rm{thr}}$, from whereon it
increases again (see the line $\theta=0$ in fig.~\ref{mandelstam_pm}).

\begin{figure}[ht]
\centerline{\resizebox{0.5\textwidth}{!}{%
 \includegraphics{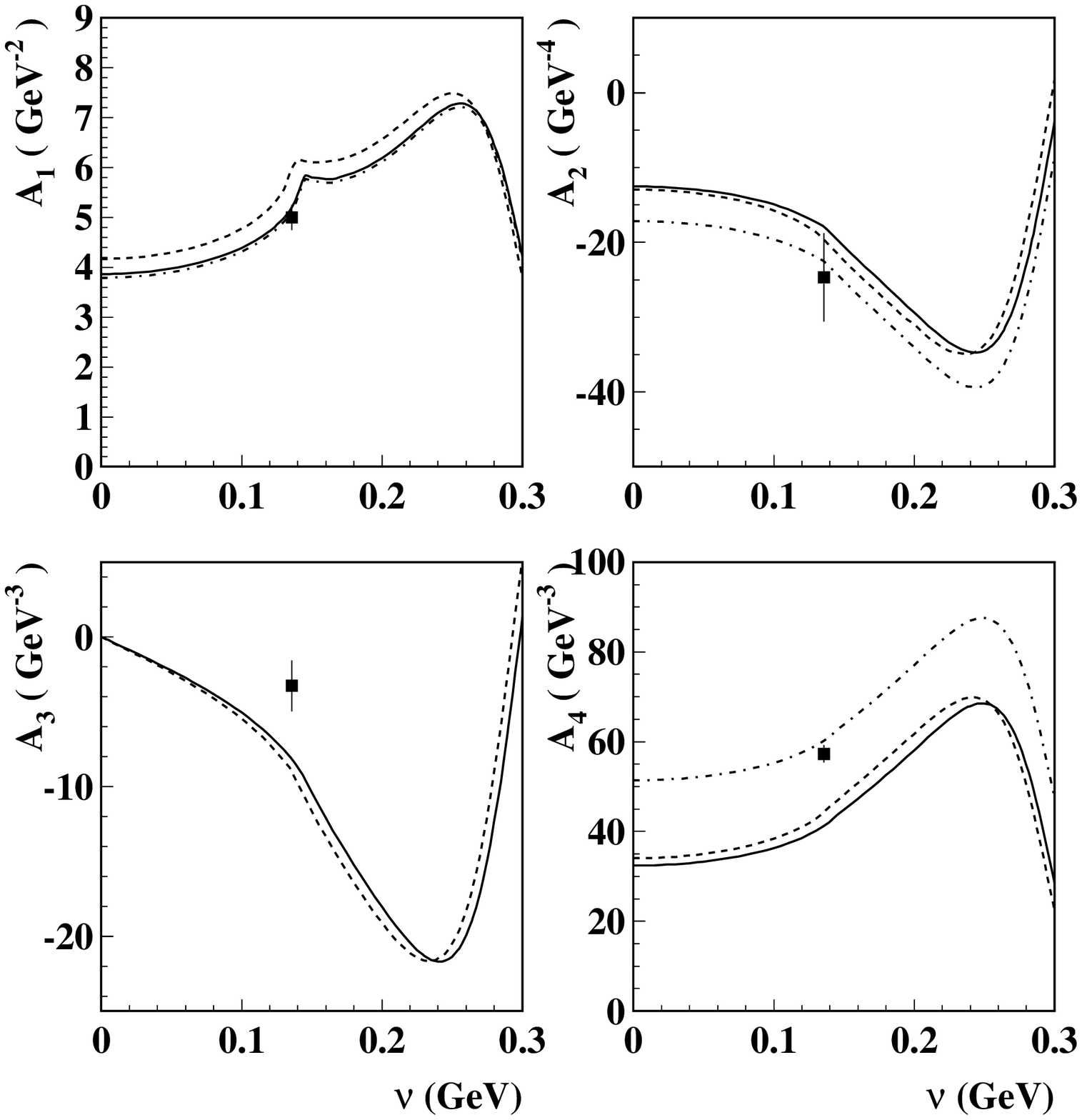}
}} \caption{ The real parts of the amplitudes $A_i^{(p\pi^0)}$ for the reaction $\gamma
p\to\pi^0 p$ as function of $\nu$ and at $t=t_{thr}$. Solid lines: dispersive
contributions according to eqs.~(\ref{eq:dr1}) and (\ref{eq:dr2}) as evaluated with the
imaginary amplitudes from MAID05 containing partial waves up to $L_{\rm{max}}=3$. Dashed
lines: same results calculated with SAID. The dashed-dotted lines are obtained by adding
the vector-meson contributions of eq.~(\ref{eq:46}) to the MAID result. The data points
near threshold are derived from the experimental values of Ref.~\cite{Sch01} for the S
and P waves plus the MAID correction for the D waves.} \label{fig:maidsaidexp}
\end{figure}

The comparison with the experiment~\cite{Sch01} in fig.~\ref{fig:maidsaidexp} shows that
the dispersion integral by itself misses the threshold data for the amplitudes $A_2-A_4$.
If we add the t-channel $\rho$ and $\omega$ poles according to MAID05, we obtain an
almost perfect agreement for $A_1$, $A_2$ and $A_4$. The apparent discrepancy between
theory and experiment for $A_3$ is an open question, which will be discussed later.

After these tests we feel safe to expand the amplitudes about the unphysical point at
$\nu=0$ and $t=M_\pi^2$ where all the involved particles are on their mass shell. For
this purpose we proceed as in our previous work~\cite{Pas05} by casting the first
amplitude in the form:
\beqn
\label{A_disp}
A_1^{\rm{disp}}(\nu,t)& = & A_1(\nu,t) - A_1^{\rm{pole}}(\nu,t) \nonumber \\
& = & \frac{eg_{\pi N}}{2M_N^2}\,\left(\kappa +\Delta_1(\nu,t)\right) \,,
\eeqn
where $\kappa$ is the anomalous magnetic moment in the respective isospin or
physical channel and $\Delta_1(\nu,t)$ the dimensionless ``FFR discrepancy''.
For convenience we also express the three other amplitudes by dimensionless
functions $\Delta_i(\nu,t)$,
\begin{eqnarray}
A_2^{\rm{disp}} (\nu,t) & = & \frac{eg_{\pi N}}{2M_N^4}\,\Delta_2(\nu,t)\,,
\nonumber \\
A_{3,4}^{\rm{disp}} (\nu,t) & = & \frac{eg_{\pi N}}{2M_N^3}\,\Delta_{3,4}(\nu,t)\,.
\label{A2_disp}
\end{eqnarray}

The functions $\Delta_i(\nu,t)$ are regular near the origin of the Mandelstam plane and
can be expanded in a power series in $\nu$ and $t$ or $\nu_B$. As is evident from
eq.~(\ref{threshold_PP}), $\nu$ is ${\mathcal{O}}(M_\pi)$ and $\nu_B$ is
${\mathcal{O}}(M_\pi^2/M_N)$ in the region of interest. Therefore the crossing-even
amplitudes $\Delta_{1,2,4}^{(+,0)}$ and $\Delta_{3}^{(-)}$ have the expansion
\beqn
\label{Delta1}
\Delta(\nu,t) & = & \delta_{00} + \delta_{20}\,\nu^2/M_\pi^2 +
\delta_{02}\,\nu_B/M_\pi
+ \delta_{40}\,\nu^4/M_\pi^4 \nonumber \\
&& + \delta_{22}\,\nu^2\nu_B/M_\pi^3 + \delta_{04}\,\nu_B^2/M_\pi^2 + \ldots
\,,
\eeqn
with the lowest expansion parameters given by
\beqn
\label{lex_parameters}
\delta_{00} & = & \Delta(0,\,M^2_\pi)\,, \nonumber\\
\delta_{20} & = &
\frac{M_\pi^2}{2}\frac{\partial^2}{\partial\,\nu^2}\,\Delta(\nu\,,M^2_\pi)\big|_{\nu=0}\,, \nonumber \\
\delta_{02} & = &
4M_N\,M_\pi\frac{\partial}{\partial\,t}\,\Delta(0,t)\big|_{t=M^2_\pi} \,. \eeqn
In the case of the crossing-odd amplitudes $\Delta_{3}^{(+,0)}$ and
$\Delta_{1,2,4}^{(-)}$ the corresponding expansion takes the form
\begin{equation}
\Delta(\nu,t) = \delta_{10}\,\nu/M_\pi + \delta_{30}\,\nu^3/M_\pi^3 +
\delta_{12}\,\nu\nu_B/M_\pi^2 + \ldots \,,
\end{equation}
with the lowest expansion parameter
\begin{equation}
\label{eq:delta10}
\delta_{10} =
M_\pi\frac{\partial}{\partial\nu}\,\Delta(\nu,M_\pi^2)\big|_{\nu=0}\,.
\end{equation}
In order to study the convergence of the multipole series involved in evaluating the
dispersion integrals of eqs.~(\ref{eq:dr1}) and (\ref{eq:dr2}) from the imaginary parts
of the amplitudes, we compare the results obtained at $t=M^2_\pi$ with values obtained at
different $t$. Table~\ref{tab1} shows our results for $\Delta_1^{(p\pi^0)}(0,t)$ for
$t=M_\pi^2$, 0, and $t_{\rm{thr}}$. The total value at $t=M^2_\pi$ corresponds to
$\delta_{00}^1=-0.078$. The slow decrease with decreasing values of $t$ is due to the
term $\delta_{02}\nu_B$.

\begin{table}
\begin{center}
\begin{tabular}{c|r@{.}lr@{.}lr@{.}l}
t & \multicolumn{2}{c}{\ \ $M_\pi^2$}
 & \multicolumn{2}{c}{$0$} &
 \multicolumn{2}{c}{$t_{\rm{thr}}$}\\
\hline
$L=0$ & $0$&$571$ & $0$&$579$ & $0$&$586$ \\
$L=1$ & $1$&$209$ & $1$&$150$ & $1$&$098$ \\
$L=2$ & $-0$&$109$ & $-0$&$089$ & $-0$&$071$ \\
$L=3$ & $0$&$044$ & $0$&$054$ & $0$&$062$ \\
\hline
sum & $1$&$715$ & $1$&$695$ & $1$&$676$ \\
\end{tabular}
\end{center}
\caption{ Contributions of the dispersion integral to $\kappa_p+\Delta_1^{(p\pi^0)}(0,t)
$ at $t=M_\pi^2$, $0$, and $t_{\rm{thr}}$ for the $p\pi^0$ channel. The rows $L=0$ to
$L=3$ indicate the contribution of the S, P, D, and F waves to the dispersion integral.
\label{tab1} }
\end{table}

The following tables~\ref{tab2} and~\ref{tab3} show the $t$ dependence of the
derivatives with regard to $\nu^2$ and $t$ at fixed $\nu=0$. The total values
at $t=M_\pi^2$ yield the constants $\delta^1_{20}$ and $\delta^1_{02}$,
respectively. The small changes with decreasing values of $t$ are due to the
higher expansion coefficients $\delta^1_{22}$ and $\delta^1_{04}$. We note that
the curvature in $\nu$ (table~\ref{tab2}) is essentially determined by the S
and P waves which add with a 1:2 ratio. However, the slope in $t$
(table~\ref{tab3}) has considerable contributions from the higher partial
waves. Although the P wave yields by far the largest contribution, the other
partial waves conspire to reduce the total slope to less than 30\% of the
P-wave result.

\begin{table}
\begin{center}
\begin{tabular}{c|r@{.}lr@{.}lr@{.}l}
t & \multicolumn{2}{c}{\ \ $M_\pi^2$}
 & \multicolumn{2}{c}{$0$} &
 \multicolumn{2}{c}{$t_{\rm{thr}}$}\\
\hline
$L=0$ & $0$&$11$ & $0$&$12$ & $0$&$12$ \\
$L=1$ & $0$&$21$ & $0$&$21$ & $0$&$22$ \\
$L=2$ & $-0$&$003$ & $-0$&$002$ & $0$&$000$ \\
$L=3$ & $-0$&$000$ & $0$&$000$ & $0$&$000$ \\
\hline
sum & $0$&$32$ & $0$&$33$ & $0$&$34$ \\
\end{tabular}
\end{center}
\caption{ Contributions of the dispersion integral to
$\frac{M_\pi^2}{2}\frac{\partial^2}{\partial\,\nu^2}\,\Delta_1^{(p\pi^0)}(\nu,t)$
for the $p\pi^0$ channel at fixed $\nu=0$. See table~\ref{tab1} for further
notation.
\label{tab2} }
\end{table}

\begin{table}
\begin{center}
\begin{tabular}{c|r@{.}lr@{.}lr@{.}l}
t & \multicolumn{2}{c}{\ \ $M_\pi^2$}
 & \multicolumn{2}{c}{$0$} &
 \multicolumn{2}{c}{$t_{\rm{thr}}$}\\
\hline
$L=0$ & $-0$&$22$ & $-0$&$23$ & $-0$&$24$ \\
$L=1$ & $1$&$62$ & $1$&$66$ & $1$&$71$ \\
$L=2$ & $-0$&$54$ & $-0$&$57$ & $-0$&$59$ \\
$L=3$ & $-0$&$29$ & $-0$&$26$ & $-0$&$24$ \\
\hline
sum & $0$&$57$ & $0$&$59$ & $0$&$63$ \\
\end{tabular}
\end{center}
\caption{ Contributions of the dispersion integral to
$4M_N\,M_\pi\frac{\partial}{\partial\,t}\,\Delta_1^{(p\pi^0)}(0,t)$ for the $p\pi^0$
channel at fixed $\nu=0$. See table~\ref{tab1} for further notation. \label{tab3}}
\end{table}

\begin{table}
\begin{center}
\begin{tabular}{c|r@{.}lr@{.}lr@{.}lr@{.}l}
 & \multicolumn{2}{c}{\ \ $\delta_{00}$}
 & \multicolumn{2}{c}{$\delta_{10}$} &
 \multicolumn{2}{c}{$\delta_{20}$}
 & \multicolumn{2}{c}{$\delta_{02}$}\\
\hline
$A_1$ & $-0$&$08\ (+0.04)$ & \multicolumn{2}{c}{\ \ -} & $0$&$32$ & $0$&$57\ (+1.11)$\\
$A_2$ & $-4$&$54\ (-1.87)$ & \multicolumn{2}{c}{\ \ -} &
$-1$&$26$ & $3$&$47\ (-1.55)$ \\
$A_3$ & \multicolumn{2}{c}{\ \ -} & $-2$&$23$ &
\multicolumn{2}{c}{\ \ -} & \multicolumn{2}{c}{\ \ -} \\
$A_4$ & $13$&$02\ (+8.17)$ & \multicolumn{2}{c}{\ \ -} &
$2$&$23$ & $-2$&$48\ (+6.98)$ 
\end{tabular}
\end{center}
\caption{  The leading expansion coefficients for the $p\pi^0$ amplitudes from
the dispersion integral (see eqs.~(\ref{A_disp})-(\ref{eq:delta10}) for definitions) and
the vector meson t-channel contributions (in brackets). \label{tab4} }
\end{table}

In table~\ref{tab4} we list the leading expansion coefficients of the 4 invariant
amplitudes for the $p\pi^0$ channel. The additional contributions of the vector meson
poles to the coefficients $\delta_{00}$ and $\delta_{02}$ are given in brackets. It is
obvious that these $t$-channel effects play an important role in neutral pion
photoproduction. The strong competition between $s$- and $t$-channel contributions to
$\delta_{00}$ and $\delta_{02}$ reflects the previous discussion on subtracted DRs.
According to eqs.~(\ref{A_disp})-(\ref{Delta1}) the subtraction functions
$A_i^I(\nu=0,t)$ are determined by the respective expansion coefficients
$\delta_{00}+\delta_{02}\nu_B/M_\pi+\cdots$, and on the other hand the DRs would sample
information at $\nu=0$ from both $t$-channel reactions (for $t>0$) and $s$-channel
reactions extrapolated into the unphysical region (for $t<0$).

The numbers in table~\ref{tab4} should be compared to an expansion of the loop plus
counter terms in covariant BChPT. Such a calculation has been performed in
Ref.~\cite{Ber05} by evaluating the third-order loop corrections and supplementing them
by a fourth-order polynomial. Since the fourth-order loop corrections are large, the
resulting power series is only indicative of the expected LECs~\cite{Ber05}. However, the
coefficients compare favorably with the LECs obtained from the earlier HBChPT
calculations~\cite{Ber96}. Including for consistency an additional factor $e$ on the RHS,
we obtain from eq.~(\ref{eq:dr1}) of Ref.~\cite{Ber05} the following coefficients from
the fourth-order polynomial contribution: $\delta_{00}^1=0$,\ $\delta_{20}^1=0.53$,
$\delta_{02}^1=3.40$, $\delta_{00}^2=-6.33$, $\delta_{10}^3=-2.58$, and
$\delta_{00}^4=22.4$. These numbers are in qualitative agreement with our results in
table~\ref{tab4}. The differences are due to
\begin{enumerate}
\item[(I)] the power series expansion of the loop corrections, which has to be
added to the polynomial,
\item[(II)] the effects of higher partial waves, particularly with regard to
the t-dependence given by $\delta_{02}$, and
\item[(III)] possible s-channel resonances above 2.2~GeV and t-channel exchange
of heavier objects, not included in the dispersive approach.
\end{enumerate}

As we have shown, the dispersion calculation including the vector meson poles is able to
reproduce the experimental threshold data except for the crossing-odd amplitude $A_3$. In
order to pin down the origin of this discrepancy, we have checked the integrands of the
dispersion integrals by repeating our calculation with the SAID~\cite{SAID} analysis.
Whereas the real parts of the MAID and SAID multipoles differ in some cases, particularly
at the higher energies, the imaginary parts generally agree well for $W\lesssim2.2$~GeV.
As shown in fig.~\ref{fig:maidsaidexp} the dispersive contributions of the two models
turn out to be quite similar. Whereas the experimental threshold values for $A_2$ and
$A_4$ can be described after adding the vector meson pole contributions, no such remedy
exists for the crossing-odd amplitude $A_3$ for which both SAID and MAID are off by a
factor of 2. We have also checked the high-energy contributions by varying the onset of
the asymptotic tail in the range 1.8~GeV$\leqslant W\leqslant 2.5$~GeV and its shape from
a simple $1/W$ dependence to various Regge prescriptions. In this way we can modify the
threshold amplitudes $A_1^{thr}$, $A_2^{thr}$, and $A_4^{thr}$ by at most 10~\% in the
$\pi^0 p$ channel. However, the asymptotic contribution to $A_3^{thr}$ reaches at most
1~\% because of the better convergence for this crossing-odd amplitude.

We have also directly estimated the high-energy tail with the parameters given by the
Regge models of the 1970's~\cite{Dev74,Bar74}, which were constructed to fit the data in
the 5-20~GeV region. Since these models include the exchange of axial vector mesons and
Regge cuts corresponding to many-particle exchange, they also contribute to the
crossing-odd amplitude $A_3$. However, the predicted strength of the high-energy
contribution to $A_3^{thr}$ turns out to be even smaller and of the order $10^{-3}$ only.
On the phenomenological level, possible candidates for axial vector meson exchange
($J^{PC}=1^{++}$) are the $a_1\ (1260)$ with quantum numbers $I^G=1^-$ and the $f_1\
(1285)$ with $I^G=0^+$. The $a_1$ has the same isospin and $G$ parity as the pion, and
therefore contributes only for charged pion production. The $f_1$, on the other hand, has
positive $G$ parity and a branching ratio of about 5~\% for the $\gamma\rho^0$ channel.
It is therefore a good candidate for $\rho$ meson photoproduction and, together with the
$a_1$, also for the Regge tail of the helicity-dependent inclusive cross sections as
measured by the Gerasimov-Drell-Hearn experiment~\cite{Bia99}. However, there appear to
be no clear candidates to solve the $A_3$ puzzle by a t-channel pole term.

Let us finally discuss the multipole content of the relativistic amplitudes and
the related error bars in the threshold region. As in our previous analysis we
subtract the nucleon pole terms, which may vary within about 2~\% depending on
the value chosen for the pion-nucleon coupling constant $g_{\pi N}$. Since the
pole term constitutes about 85~\% of the total threshold amplitudes $A_2^{thr}$
and $A_3^{thr}$, the choice of $g_{\pi N}$ leads to an error of about 12~\% in
the remaining dispersive amplitudes. For $A_1^{thr}$ and $A_4^{thr}$, however,
the pole contributions are small and therefore the model error of $g_{\pi N}$
can be neglected with regard to the dispersive amplitudes. In table~\ref{tab5}
the threshold amplitudes are constructed from the experimental values of the S-
and P-wave multipoles and the MAID05 value for the D waves. The error given in
the table is obtained by adding the experimental errors for the threshold
multipoles. We recall at this point that here and in the following all values
refer to the dispersive contributions only.

As is evident from table~\ref{tab5}, $A_1^{thr}$ is dominated by the S-wave, because the
magnetic contributions cancel nearly completely. The large S-wave contribution originates
from the FFR current~\cite{Pas05} and rescattering corrections, which are of course
included in BChPT by the chirally invariant pion-nucleon coupling and the pion loops
leading to the pronounced cusp effect at the $n\pi^+$ threshold. These effects are nicely
reproduced by the dispersion integral, and the small t-channel pole contribution is well
within the experimental error bars. More precisely, the vector mesons produce large
effects of about equal size for both magnetic multipoles, but the discussed cancellation
between $M_{1+}$ and $M_{1-}$ leads to a reduction by a large factor.

Quite different physical information is sampled by $A_2^{thr}$. Due to the structure of
the associated four-vector $M_2^{\mu}$, the spin $J=1/2$ multipoles $E_{0+}$ and $M_{1-}$
do not appear in this amplitude. It is dominated by $M_{1+}$ but also receives
surprisingly large contribution from the D waves whose threshold values are not very well
known. In HBChPT~\cite{Ber96} about 90~\% of this amplitude are obtained from counter
terms.

In the case of $A_3^{thr}$ we again find a large cancellation of the magnetic multipole
contributions, which is required because the $\rho$ and $\omega$ pole terms have to
cancel exactly for symmetry reasons. However, the dispersive contributions of $M_{1+}$
and $M_{1-}$ are now large compared to $E_{0+}$, and this cancellation leads to a large
error bar. The amplitude $A_3$ shows a weak cusp effect, in accordance with the result of
HBChPT~\cite{Ber96}.

Finally, the amplitude $A_4^{thr}$ is determined by the $\Delta(1232)$ multipole $M_{1+}$
and a remarkably strong Roper multipole $M_{1-}$. Since no cancellation appears among the
dispersive contributions, the error bar of this amplitude is small. Similarly as in the
case of $A_2^{thr}$, the amplitude $A_4^{thr}$ is almost totally determined by counter
terms in both HBChPT~\cite{Ber96} and covariant BChPT~\cite{Ber05}.

\begin{table}
\begin{center}
\begin{tabular}{l|ccccc|c}
 & $E_{0+}$ &$E_{1+}$ &$M_{1+}$ &$M_{1-}$ &$D$ &total \\
\hline
$A_1$ & $4.75$ & $ 0.08$ & $  2.02$ & $-1.94$ & $ 0.08$ & $ 5.00\pm0.25$\\
$A_2$ &   $0$  & $-0.34$ & $-29.44$ &  $0$    & $ 5.13$ & $-24.7\pm5.9 $\\
$A_3$ & $2.36$ & $-0.65$ & $-18.67$ & $14.34$ & $-0.65$ & $ -3.3\pm1.7 $\\
$A_4$ & $2.36$ & $ 0.04$ & $ 40.55$ & $14.34$ & $ 0.04$ & $ 57.3\pm1.9 $\\
\end{tabular}
\end{center}
\caption{ The multipole decomposition of the dispersive part of the
experimental threshold amplitudes, constructed from the data of Ref.~\cite{Sch01} and the
MAID05 value for the D-state contributions. The errors include statistical and systematic
errors from the experiment and model errors due to uncertainties from the pion-nucleon
coupling and from the D waves, all added in quadrature. \label{tab5} }
\end{table}

\section{Summary and Outlook~\label{sec:sum}}

We have studied the photoproduction of pions by means of dispersion relations
at $t=const$ for the relativistic amplitudes $A_1$ to $A_4$, using the
imaginary parts of the amplitudes as input to evaluate the real parts. The
calculations are performed with and compared to the most recent version MAID05,
and it is our final goal to put this analysis on the basis of dispersion
theory. Along these lines the present exploratory work on neutral pion
photoproduction has the aim to (I) test the numerical procedure and the data
basis by trying to reproduce the precision data near threshold and (II) predict
the low-energy constants of baryon chiral perturbation theory (BChPT) by global
properties of the excitation spectrum.

After subtraction of the nucleon and pion pole terms, the remaining
``dispersive'' amplitudes are regular functions of the Mandelstam variables in
the subthreshold region. In particular they may be expanded in a power series
in the two independent variables $\nu$ and $t$ about the point $\nu=0$,
$t=M_\pi^2$. This series converges in a circle with a radius determined by the
threshold of pion production. The singularity at threshold is, of course, well
described by the loop corrections of BChPT, which also can be expanded in a
power series below threshold. The difference of the expansions of the full
amplitude and the loop contribution is described by the low-energy constants
necessary to apply BChPT to the data analysis.

We find that the threshold amplitude $A_1^{thr}$ is well described by the dispersion
integral within the experimental error. As in our previous work we also obtain a good
agreement with the FFR sum rule that connects the amplitude $A_1$ to the anomalous
magnetic moment of the nucleon. In the case of $A_2^{thr}$ and $A_4^{thr}$, however, we
have to include additional contributions of t-channel vector meson exchange. The
importance of such effects for neutral pion photoproduction has been noted long ago. As
was to be expected, our calculations also show that these fixed poles at $t=m_V^2$ can
not be obtained from the dispersion integrals. Moreover, though these mesons play an
important role for the unitarization process at the higher energies, their global effect
on the dispersion integrals for the threshold amplitudes is surprisingly small. If we add
the vector meson pole terms to the dispersion integrals, we obtain a considerable
improvement also for $A_2^{thr}$ and $A_4^{thr}$, and a perfect fit would require only a
mo\-dest change of the coupling constants with respect to the MAID values.

However, the predicted value for $A_3^{thr}$ is at variance with the data by
more than 2 standard deviations. This is a serious problem for the crossing-odd
amplitude $A_3$, because a t-channel pole contribution requires the exchange of
an axial vector meson. Unfortunately, the known axial vector mesons have either
the wrong isospin or the wrong G parity for neutral pion photoproduction.

In view of the apparent discrepancy for $A_3^{thr}$ we have carefully checked
all the ingredients of our calculation. We have repeated the integrations using
the absorptive amplitudes given by SAID~\cite{SAID}. In spite of occasional
differences between the MAID and SAID multipole analyses, particularly at
energies $W\gtrsim2$~GeV, both models predict quite similar amplitudes up to
the $\Delta(1232)$ resonance. We have further studied the influence of the
high-energy region by varying the upper limit of integration in the range of
1.8~GeV$\leqslant W\leqslant 2.5$~GeV and by fitting the high-energy tail with
various shapes from a simple $1/W$ behavior to several Regge forms. In this way
the threshold amplitudes change by about 5~\% for $A_1^{thr}$ and a few per
cent for $A_2^{thr}$ and $A_4^{thr}$, whereas the possible error for
$A_3^{thr}$ is less than 1~\%. If we directly apply various Regge models,
including phenomenological Regge poles and cuts, the high-energy contribution
turns out to be even smaller, particularly in the case of $A_3^{thr}$. Even
independent of any model assumption, the crossing-odd amplitude $A_3$ has the
best convergence property of all the amplitudes, and therefore an extremely
strong asymptotic contribution would be necessary to change $A_3^{thr}$ by a
factor of 2.

As we have shown, the experimental error for $A_3^{thr}$ is very large due to a
near total cancellation between the leading magnetic multipoles. For this
reason the D state contribution may give rise to some concern. Its contribution
has been estimated to account for about 20~\% of $A_3^{thr}$, which assumption
is based on a reasonable extrapolation to the threshold region. However, local
fits in certain energy bins often yield large fluctuations of the D-state
background. Therefore a large model error for this background can not be
excluded and may be partially responsible for the $A_3$ problem.

With all these caveats in mind we have evaluated the dispersive and vector meson
contributions of the relativistic amplitudes and expanded the result in a power series in
$\nu$ and $t$ as discussed in detail in the previous section. The results nicely compare
with the low-energy constants derived from a fit of covariant BChPT to the threshold
data. Except for the $A_3$ problem, the differences between the two approaches are due to
additional loop terms. In view of the unexpectedly large loop corrections at
$4^{\rm{th}}$ order HBChPT, an extension of the existing $3^{\rm{rd}}$ order covariant
BChPT to the next order will be of great general interest.

On the side of the dispersive approach, the simple addition of the vector meson
pole terms to the dispersive amplitude is of course only justified near
threshold where the scattering phases are small. In general it will be
necessary to unitarize the full amplitude comprising the nucleon, pion, and
vector meson pole terms as well as the complex contribution of the continuum.
This will require an iterative procedure including modifications of the model
parameters encoded in the helicity amplitudes of the resonances. A further
challenge is to describe the gradual transition from vector meson poles near
threshold to Regge propagators at large energies, without introducing spurious
singularities and respecting the crossing symmetry of the invariant amplitudes.

In view of the large D-wave background and the delicate cancellation of various
multipoles in $A_3^{thr}$, dedicated experiments to measure the D-wave
contribution in the region between threshold and the $\Delta(1232)$ resonance
would be extremely helpful. Whereas the  difference of recoil and target
polarization in forward direction is directly proportional to the amplitude
$A_3$ in the high-energy limit, this observable does not appear to be very
sensitive to $A_3$ in the threshold region. We are presently studying several
double-polarization observables with regard to an enhanced sensitivity for
$A_3$ and the D-wave background.

\section*{Acknowledgements}

This work was supported by the Deutsche Forschungsgemeinschaft (SFB 443) and the EU
Integrated Infrastructure Initiative Hadron Physics Project under contract number
RII3-CT-2004-506078.

%

%

\begin{figure}[ht]
\centerline{\resizebox{0.5\textwidth}{!}{%
  \includegraphics{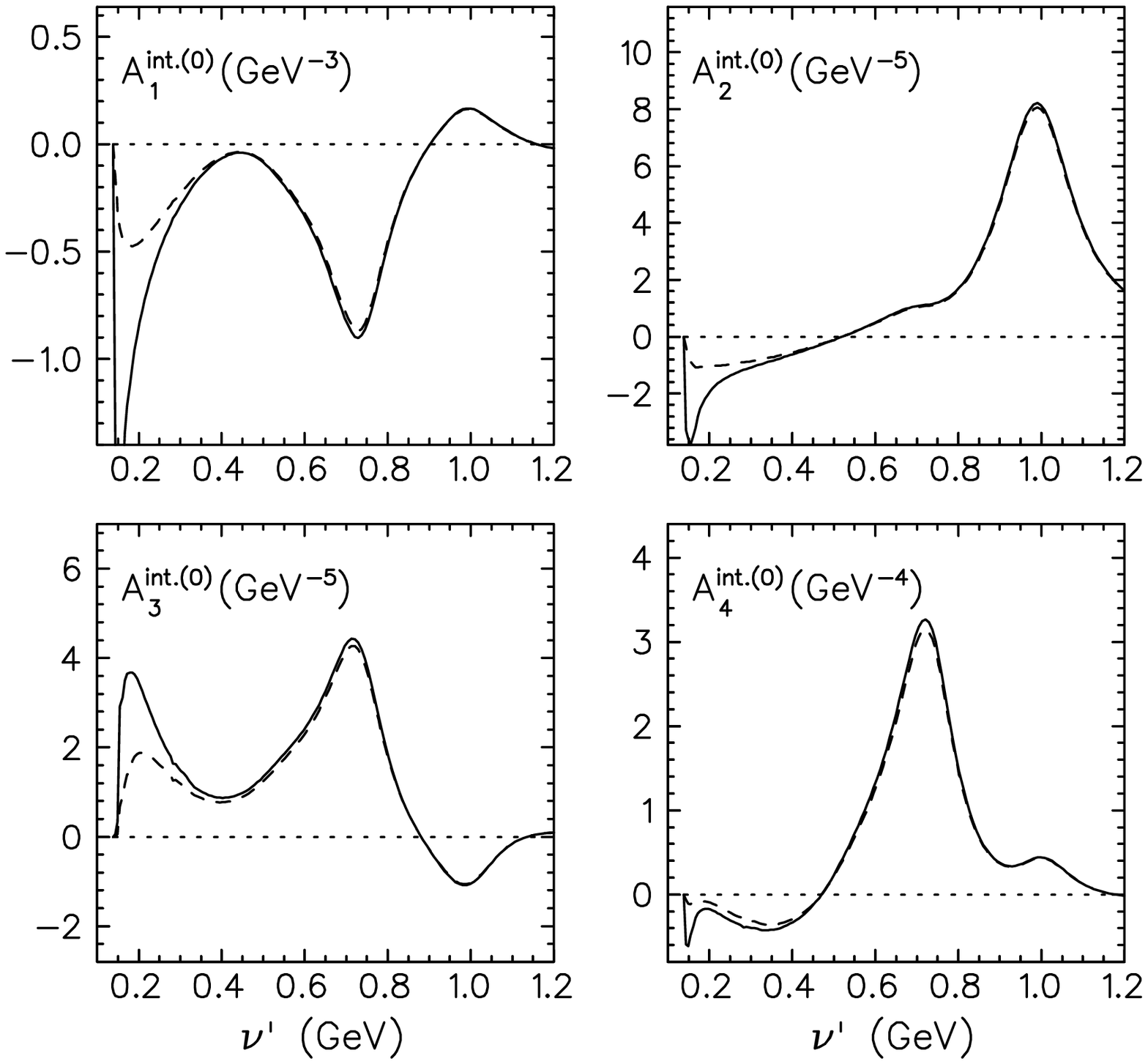}
}} \caption{ The integrands of eqs.~(\ref{eq:dr1}) and (\ref{eq:dr2}) for the amplitudes
$A_i^{(0)}$ obtained with $\nu=\nu_{\rm{thr}}$ (solid lines) and $\nu=0$ (dashed lines).}
\label{fig:integrandszero}
\end{figure}
%

%

%

\begin{thebibliography}{99}

\bibitem{Pas05}
B. Pasquini, D. Drechsel, and L. Tiator, Eur. Phys. J. A 23 (2005) 279.

\bibitem{Fub66}
S. Fubini, G. Furlan, and C. Rossetti, Nuovo Cim. 40 (1965) 1171.

\bibitem{MAID}
D. Drechsel, O. Hanstein, S.S. Kamalov, and L. Tiator, Nucl. Phys. A 645 (1999)
145; http://www.kph.uni-mainz.de/MAID/.

\bibitem{Bec90}
R. Beck et al., Phys. Rev. Lett. 65 (1990) 1841.

\bibitem{Ber91}
V. Bernard, N. Kaiser, J. Gasser, and U.-G. Mei{\ss}ner, Phys. Lett. B 268
(1991) 291; V. Bernard, N. Kaiser, and U.-G. Mei{\ss}ner, Nucl. Phys. B 383
(1992) 442.

\bibitem{Bec99}
T. Becher and H. Leutwyler, Eur. Phys. C 9 (1999) 643; B. Kubis and U.-G.
Mei{\ss}ner, Nucl. Phys. A 679 (2001) 698; T. Fuchs, J. Gegelia, G. Japaridze,
and S. Scherer, Phys. Rev. D 68 (2003) 056005.

\bibitem{Ber05}
V. Bernard, B. Kubis, and U.-G. Mei{\ss}ner, Eur. Phys. J. A 25 (2005) 419.

\bibitem{Che57}
G.F. Chew et al., Phys. Rev. 106 (1957) 1345.

\bibitem{Han98}
O. Hanstein, D. Drechsel, and L. Tiator, Nucl. Phys. A 632 (1998) 561.

\bibitem{Bjo65} J.D. Bjorken and S.D. Drell, {\it Relativistic Quantum Fields},
McGraw-Hill, New York (1965).

\bibitem{Man58}
S. Mandelstam, Phys. Rev. 112 (1958) 1344; 115 (1959) 1741; 115 (1959) 1752.

\bibitem{Fra60}
See, e.g., S.C. Frautschi and J.D. Walecka, Phys. Rev. 120 (1960) 1486;
W.R.~Frazer and J.R.~Fulco, Phys. Rev. 119 (1960) 1420.

\bibitem{Bal61}
J.S. Ball, Phys. Rev. 124 (1961) 2014.

\bibitem{Whi58}
E.T. Whittaker and G.N. Watson, {\it Modern Analysis}, p.~322, Cambridge
University Press (1958).

\bibitem{Oeh59}
R. Oehme and J.G. Taylor, Phys. Rev. 113 (1959) 371.

\bibitem{Geh69}
G. v. Gehlen, Nucl. Phys. B 9 (1969) 17.

\bibitem{DDBPMV}
D.~Drechsel, B.~Pasquini and M.~Vanderhaeghen, Phys. Rep. {378} (2003) 99.

\bibitem{Gui97}
M. Guidal, J.-M. Laget, and M. Vanderhaeghen, Nucl. Phys. A 627 (1997) 645 and
Phys. Lett. B 400 (1997) 6.

\bibitem{Azn03}
I.G. Aznauryan, Phys. Rev. C 67 (2003) 015209.

\bibitem{Dev74}
R.C.E. Devenish, D.H. Lyth, and W.A. Rankin, Phys. Lett. 52B (1974) 227.

\bibitem{Bar74}
I.S. Barker, A. Donnachie, and J.K. Storrow, Nucl. Phys. B 79 (1974) 431.

\bibitem{Maid01}
S.~S.~Kamalov, S.~N.~Yang, D.~Drechsel, O.~Hanstein and L.~Tiator,
Phys. Rev. C {64} (2001) 032201.

\bibitem{PDG}
S. Eidelman et al., Phys. Lett. B 592 (2004) 1.

\bibitem{Tia03}
L.~Tiator, D.~Drechsel, S.S.~Kamalov, and S.N.~Yang,
Eur. Phys. J. A {17} (2003) 357.

\bibitem{Tia04}
L.~Tiator, D.~Drechsel, S.S.~Kamalov, M.M.~Giannini, E.~Santopinto, and
A.~Vassallo, Eur. Phys. J. A {19} (2004) 55.

\bibitem{DrMaid02}
S.~S.~Kamalov, L.~Tiator, D.~Drechsel, R.~A.~Arndt, C.~Bennhold, I.~I.~Strakovsky and R.~L.~Workman,
Phys. Rev. C {66} (2002) 065206.

\bibitem{Sch01}
A. Schmidt et al., Phys. Rev. Lett. 87 (2001) 232501.

\bibitem{Ber96}
V. Bernard, N. Kaiser, and U.-G. Mei{\ss}ner, Z. Phys. C 70 (1996) 483; V.
Bernard, N. Kaiser, and U.-G. Mei{\ss}ner, Phys. Lett. B 378 (1996) 337; V.
Bernard, N. Kaiser, and U.-G. Mei{\ss}ner, Eur. Phys. J. A 11 (2001) 209.

\bibitem{SAID}
R.A. Arndt, W.J. Briscoe, I.I.Strakovsky, and R.L. Workman, Phys. Rev. C 66
(2002) 055213; http://gwdac.phys.gwu.edu/.

\bibitem{Bia99}
N. Bianchi and E. Thomas, Phys. Lett. B 450 (1999) 439.

\end{thebibliography}
\end{document}